\documentclass[10pt,letterpaper,twocolumn]{article} 

\usepackage{ol2}
\usepackage[draft]{hyperref}
\usepackage{amsmath}

\begin{document}

\twocolumn[ 

\title{Probing one-dimensional topological phases in waveguide lattices with broken chiral symmetry}


\author{Stefano Longhi}

\address{Dipartimento di Fisica, Politecnico di Milano and Istituto di Fotonica e Nanotecnologie del Consiglio Nazionale delle Ricerche, Piazza L. da Vinci 32, I-20133 Milano, Italy (stefano.longhi@polimi.it)}

\begin{abstract}
One-dimensional lattices with chiral symmetry are known to possess quantized Zak phase and nontrivial topological phases. Here it is shown that quantized Zak phase and nontrivial
 edge states, partially protected by inversion symmetry rather than chiral symmetry, can be observed and probed in the bulk exploiting continuous-time photonic quantum walk in zig-zag waveguide arrays. Averaged beam displacement measurements can detect quantized Zak phase and non-trivial topological phases in the extended  Su-Schrieffer-Heeger model with broken chiral symmetry. 
\end{abstract}

 ] 
 
{\em Introduction.} Topological states of matter are known to exhibit protected edge states and
robust quantized features in their bulk \cite{r1}. In the past two decades, there has been a surge of interest
in studying topological orders in non-electronic systems with a high degree of controllability,
such as in cold atoms, trapped ions and photons. In particular, topological photonics has emerged as a rapidly area of research \cite{r2,r3}. 
Topological phases are parametrized by integer topological
invariants, which can be probed at the surface (owing to the bulk-boundary
correspondence \cite{r1}) or directly in the bulk.
 Several photonic simulators, such as those based on discrete or continuous-time quantum walks, have been
 used to probe topological invariants both at the edges and in the bulk \cite{r3bis,r4,r5,r6,r7,r8,r9,r10,r11,r12,r13}, with the ability to test topological features even for non-Hermitian models \cite{r9,r10,r14,r14bis,r15,r16,r16bis,r17}.
 In one-dimensional (1D) systems with
chiral symmetry, such as in the
Su-Schrieffer-Heeger (SSH) model \cite{r1}, the bulk topological properties of the Bloch bands
are characterized by the winding number and quantized Zak phase \cite{r1,r19,r20}. Since the Zak phase is the Berry phase acquired by a Bloch eigenstate
during an adiabatic evolution across the whole Brillouin zone \cite{r19}, its measure requires rather generally to apply some external force and to extract the geometric phase contribution
after a closed adiabatic loop \cite{r20,r21,r22}. In a recent work  \cite{r13} a very simple method has been suggested and experimentally demonstrated for probing the Zak phase and topological invariants in the bulk  of 1D chiral systems, based on  discrete-time quantum walks of twisted light. The main result is that, for chiral systems, the mean displacement of the walker, when the initial wavepacket is localized on a
single site, provides asymptotically a direct measure of the winding number \cite{r13}. A similar method for a non-Hermitian version of the SSH model was earlier suggested in\cite{r23} and demonstrated in \cite{r9}.\\
Chiral symmetry, generally required in Altland-Zirnbauer classification for 1D systems \cite{uff}, ensures Zak phase quantization and the occurrence of topological non-trivial edge states. However, recent works have suggested and experimentally demonstrated that quantization of Zak phase and non-trivial edge states, partially protected by inversion symmetry, can be observed in 1D lattices with broken chiral symmetry \cite{r24,r26}.\\
In this Letter it is shown that spatial beam shift observed in photonic quantum walks in waveguide lattices can provide a simple and experimentally accessible tool for exploring quantized Zak phase and non-trivial edge states  in 1D insulating systems  with broken chiral symmetry. Spatial beam displacements have been often related to topological effects and geometric Berry phase. For example, in the photonic spin Hall effect the topological coupling between the spin and the trajectory of a light beam can lead to a transverse shift in polarization components for reflected or transmitted beams at an interface \cite{referee}. Here, it is shown rather generally that in a two-band 1D insulating system the time-average beam displacement of photons, under initial single-site excitation, provides a measure of Berry phase when the phase is quantized, i.e. when the system possesses non-trivial topological states. As an example, a simple waveguide array design is suggest that realizes the extended SSH model with next nearest neighbor couplings \cite{r26,r27,r28}, where long-range hopping breaks chiral symmetry but inversion symmetry ensures partial topological protection of degenerate edge states.\par
{\it Quantization of Zak phase in 1D two-band systems.}  
We consider a 1D two-band  lattice, each unit cell hosting two sites, one on sublattice A, and one on sublattice B. The single-particle Hamiltonian in bulk momentum space representation reads \cite{r1,r3}
\begin{equation}
\mathcal{H}(k)= 
\left(
\begin{array}{cc}
h_A(k) & h_1(k) \\
h_1^*(k) & h_B(k)
\end{array}
\right)= \frac{h_A(k)+h_B(k)}{2} \sigma_0 + \boldsymbol{a}(k) \cdot \boldsymbol{\sigma}
\end{equation}
In the above equations, $h_A(k)$ and $h_B(k)$ account for hopping among sites in each of sublattices A and B, respectively, $h_1(k) \equiv H(k) \exp[i \varphi(k)]$ describes cross-hopping among sites of different sublattices, $k$ is the quasi momentum ($-\pi \leq k < \pi$), $\sigma_0$ is the $2 \times 2$ identity matrix, $\boldsymbol{\sigma}=(\sigma_x,\sigma_y,\sigma_z)$ describes the vector of Pauli matrices, and
\begin{equation}
\boldsymbol{a}(k)=\left( H \cos \varphi, -H \sin \varphi, \frac{h_A-h_B}{2} \right).
\end{equation}

 For Hermitian lattices without gauge fields, $h_{A,B}(k)$ are real with $h_{A,B}(-k)=h_{A,B}(k)$, $H(-k)=H(k)$ and $\varphi(-k)=-\varphi(k)$. For example, for the SSH model with staggered nearest neighbor hopping $t_1$, $t_2$ and next-nearest neighbor hopping $t_{A,B}$ [Fig.1(a)], one has
\begin{equation}
h_{A,B}(k)= \delta_{A,B}+2 t_{A,B} \cos (k) \; , \; \; h_1(k)=t_2+t_1 \exp(-ik)
\end{equation}
  \begin{figure}[htb]
\centerline{\includegraphics[width=8.4cm]{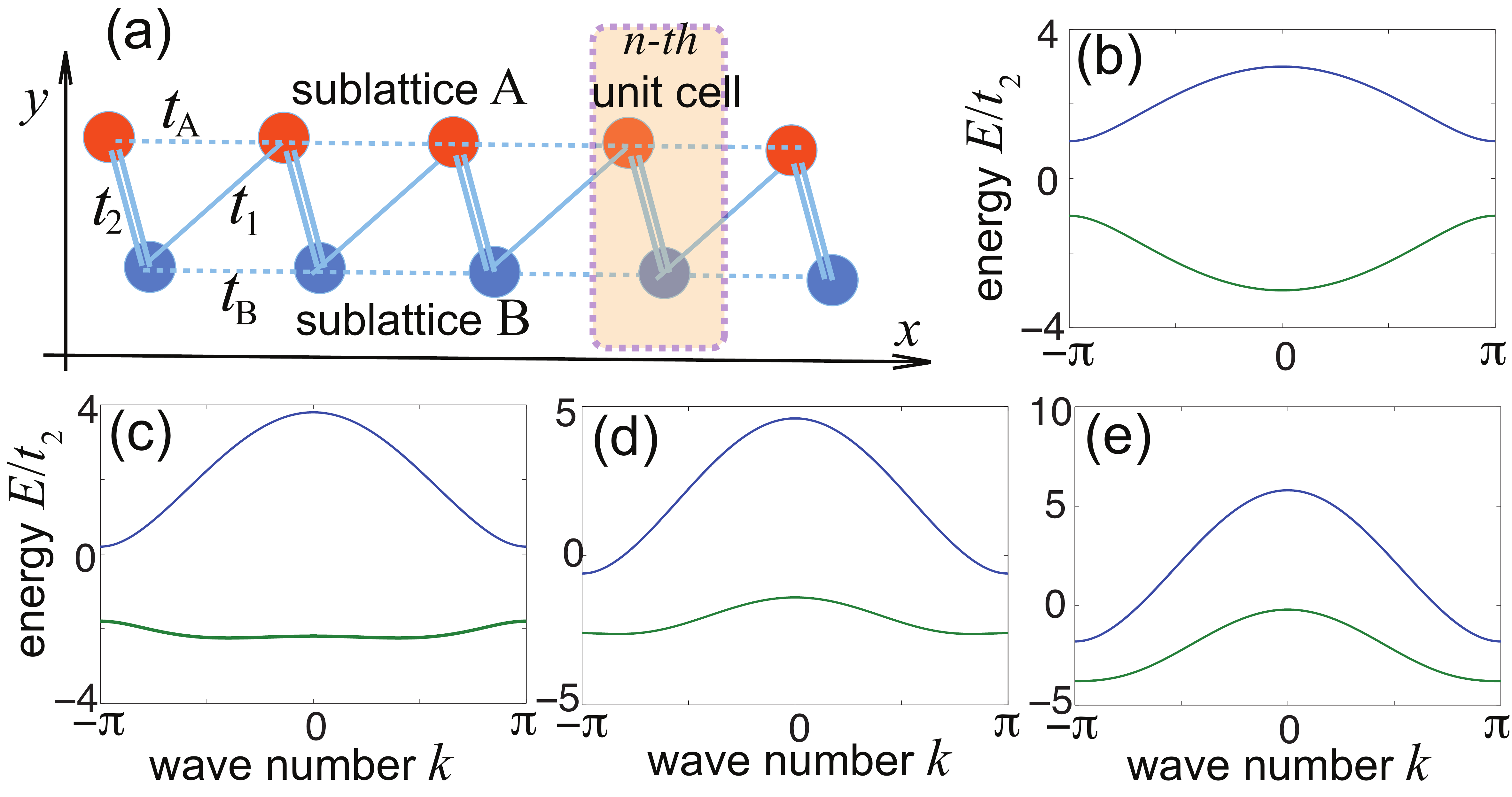}} \caption{ \small
(Color online) (a) Schematic of a zig-zag binary waveguide lattice with staggered nearest ($t_1, \; t_2$) and next-nearest neighbor ($t_A, \; t_B$) hopping. (b-e) Energy spectrum (dispersion curves $E_{\pm}(k)$ versus Bloch wave number $k$) for the extended SSH model with inversion symmetry ($t_A=t_B$, $\delta_A=\delta_B=0$) for $t_1 /t_2=2$ and for increasing values of $t_A/t_2$: (b)  $t_A/t_2=0$ (ordinary SSH model), (c) $t_A/t_2=0.4$, (d) $t_A/t_2=0.8$, and (e) $t_A/t_2=1.4$. In (b) and (c) the gap is open and the bandgap is direct ($t_A<t_2/2$); in (d) the gap is still open but the bandgap is indirect ($t_2/2<t_A<t_1/2$); in (e) the gap is closed ($t_A> t_1/2$).}
\end{figure} 
where $\delta_{A,B}=\pm \delta$ describes the energy offset of sites A and B, respectively. Chiral ($\mathcal{C}$) and inversion ($\mathcal{R}$) symmetries play a major role in determining the nontrivial topological properties of the Hamiltonian (1). $\mathcal{C}$ and $\mathcal{R}$ are defined as $k$-independent unitary operators such that $\mathcal{C}\mathcal{H}(k) \mathcal{C}^{-1}=-\mathcal{H}(k)$ and $\mathcal{R} \mathcal{H}(k) \mathcal{R}^{-1}=\mathcal{H}(-k)$, with $\mathcal{C}^2=\mathcal{R}^2=\sigma_0$. Chiral (or sublattice) symmetry, with $\mathcal{C}=\sigma_z$, necessarily requires that $h_A(k)=h_B(k)=0$, whereas inversion symmetry, with $\mathcal{R}= \sigma_x$, is ensured under the less stringent requirement $h_A(k)=h_B(k) \neq 0$.  For example, the standard SSH with nearest neighbor hopping ($t_A=t_B=0=\delta=0$) shows both chiral and inversion symmetry \cite{r1}, while a non vanishing next- nearest neighbor hopping $t_{A,B}$ trivially breaks chiral symmetry but not the inversion symmetry provided that $\delta=0$ and $t_A=t_B$ \cite{r24,r26}.\\
 \begin{figure}[htb]
\centerline{\includegraphics[width=8.6cm]{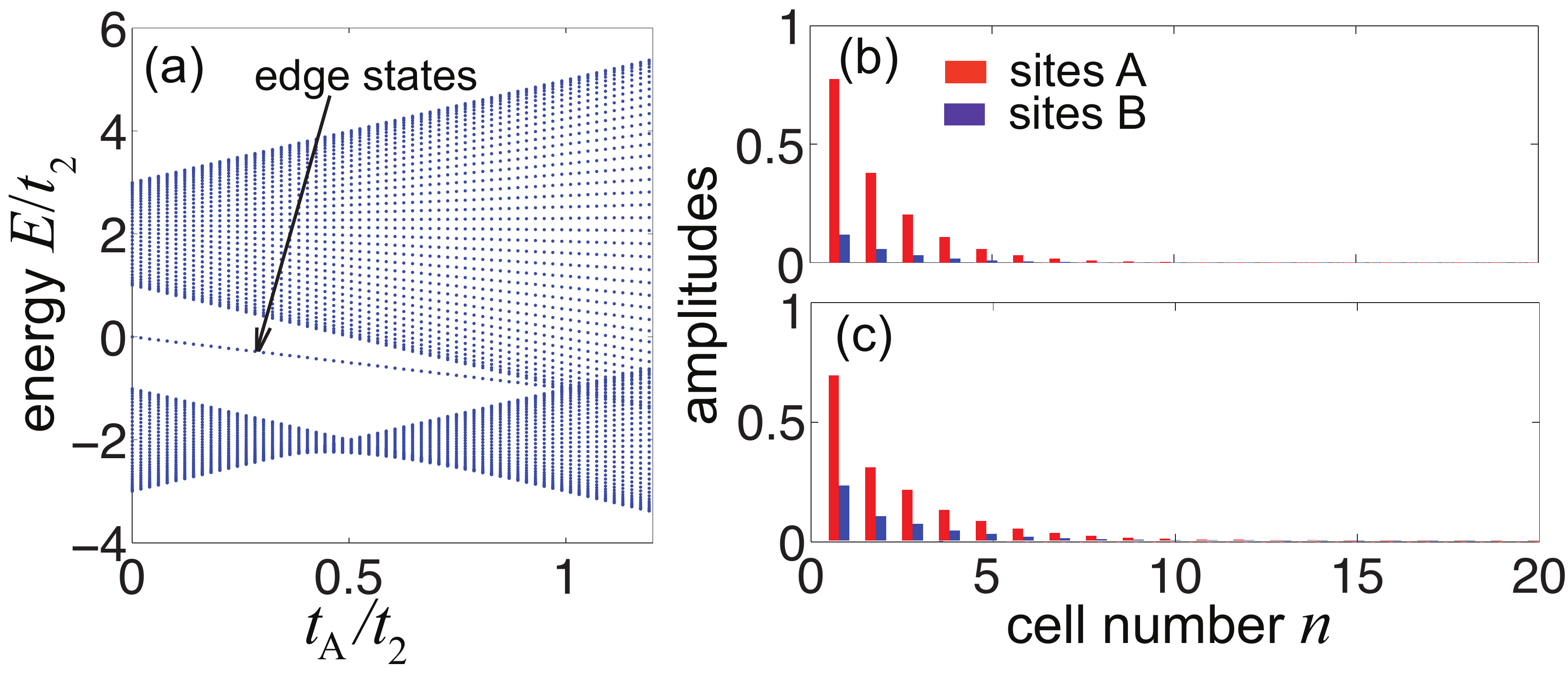}} \caption{ \small
(Color online) (a) Energy spectrum versus the ratio $t_A/t_2$ for the extended SSH Hamiltonian with inversion symmetry ($\delta_A=\delta_B=0$, $t_A=t_B$) for the non-trivial topological phase $t_1/t_2=2$. The chain comprises $N=40$ unit cells. The dotted curve in the gapped region corresponds to the energy $E_{edge}$ of the two nearly-degenerate edge states. (b) and (c) show the detailed distributions of amplitudes $|a_n|$ and $|b_n|$ is the sublattices A and B for the left edge state (the right edge state is simply obtained by the inversion symmetry). In (b) $t_A/t_2=0.3$, in (c) $t_A/t_2=0.6$. Note that the energy of the edge states deviates from the midgap energy level and both sublattices A and B are occupied, indicating sublattice symmetry breaking.}
\end{figure} 
The energy dispersion curves $E_{\pm}(k)$ and eigenfunctions  $\mathbf{u}_{\pm}(k)$ of the two lattice bands are given by
\begin{eqnarray}
E_{\pm}(k)  =  \frac{h_A(k)+h_B(k)}{2} \pm \sqrt{H^2(k)+ \epsilon^2(k)} \;\;\;\;\;\;\;\; \\
\mathbf{u}_+  =  \left( 
\begin{array}{c}
\cos \frac{\theta}{2} \\
\sin \frac{\theta}{2} \exp(-i \varphi)
\end{array}
\right) , 
\mathbf{u}_-  =  \left( 
\begin{array}{c}
\sin \frac{\theta}{2} \\
- \cos \frac{\theta}{2} \exp(-i \varphi)
\end{array}
\right) 
\end{eqnarray}  
where we have set $\epsilon(k) \equiv [h_A(k)-h_B(k)]/2$ and where the angle $\theta=\theta(k)$ is defined via the relation 
\begin{equation}
\tan \theta(k)=H(k) / \epsilon(k).
\end{equation}
 The Zak phase \cite{r19} for the two lattice bands can be readily calculated from the relation 
 \begin{equation}
\gamma_{\pm}=i \int_{-\pi}^{\pi} dk \langle \mathbf{u}_{\pm} | \frac{\partial \mathbf{u}_{\pm}}{\partial k} \rangle
\end{equation}  
and reads explicitly
\begin{equation}
\gamma_+=\int_{-\pi}^{\pi} dk \frac{\partial \varphi}{\partial k} \sin^2 \frac{\theta}{2} \; , \; \gamma_-=\int_{-\pi}^{\pi} dk \frac{\partial \varphi}{\partial k} \cos^2 \frac{\theta}{2}
\end{equation}
Equation (8) indicates that the Zak phase, rather generally, is not quantized and different for the two bands.  However, if the inversion symmetry is not broken, i.e. for $h_A(k)=h_B(k)$, from Eq.(6) one has $\theta(k)=\pi/2$ independent of $k$ and thus 
\begin{equation}
\gamma_{+}=\gamma_{-}= \frac{1}{2} \int_{-\pi}^{\pi} dk \frac{\partial \varphi}{\partial k}= \pi \mathcal{W}
\end{equation}
where $\mathcal{W} \equiv (1/ 2 \pi) \oint (a_y da_x-a_x da_y)/ (a_x^2+a_y^2)$ is the winding number of the Hamiltonian (1). Equation (9) shows that the Zak phase is quantized even when the chiral symmetry is explicitly broken, provided that the inversion symmetry is conserved. In fact, a non-vanishing term $h_A=h_B$ in the Hamiltonian changes the energy spectrum but not the eigenfunctions $\mathbf{u}_{\pm}$, so that only the dynamical phase (but not the Berry phase) is changed. \par
{\em Zak phase and wave packet displacement.} 
Zak phase measurements usually  require to extract the geometric phase contribution when Bloch oscillations are induced in the system \cite{r20}, which is a nontrivial task. Can we probe the quantized Berry phase in the bulk without resorting to Bloch oscillations? For a chiral Hamiltonian, it has been recently shown that in discrete-time quantum walks of twisted photons the mean chiral displacement of a freely evolving  wavepacket can be used to probe the Zak phase \cite{r13}. Here we show that  spatial beam displacement in continuous-time quantum walks can  reveal Zak phase in the less stringent case of Hamiltonians with inversion symmetry but broken chiral symmetry. Let us indicate by $a_n(t)$ and $b_n(t)$ the occupation amplitudes of sublattices A and B in the $n$-th cell, and let us assume single-site excitation at initial time, for example excitation of sublattice A, $a_n(0)=\delta_{n,0}$ and $b_n(0)=0$. We define the mean wave packet displacement at successive time $t$ as $\langle n(t) \rangle= \sum_n n |a_n(t)|^2+\sum_n n |b_n(t)|^2$. After some lengthy but straightforward calculations, one can show that for the general Hamiltonian (1) one has $\sum_n n |a_n(t)|^2=0$ and 
\begin{equation}
\langle n(t) \rangle= 	\sum_n n |b_n|^2=\frac{1}{2 \pi} \int_{-\pi}^{\pi} dk \frac{\partial \varphi}{\partial k}\sin^2 [\Delta(k)t] \sin^2 \theta(k) 
\end{equation}
where we have set $\Delta(k) \equiv \sqrt{H^2(k)+\epsilon^2(k)}$. For a gapped Hamiltonian, i.e. provided that $\Delta(k)$ does not vanish as $k$ spans the entire Brillouin zone, the time average of the wave packet displacement reads
\begin{equation}
\overline{\langle n(t) \rangle} \equiv \frac{1}{T} \int_0^T dt \langle n(t) \rangle \simeq \frac{1}{4 \pi} \int_{-\pi}^{\pi} dk \frac{\partial \varphi}{\partial k} \sin^2 \theta(k)
\end{equation}
\begin{figure*}[htb]
\centerline{\includegraphics[width=17cm]{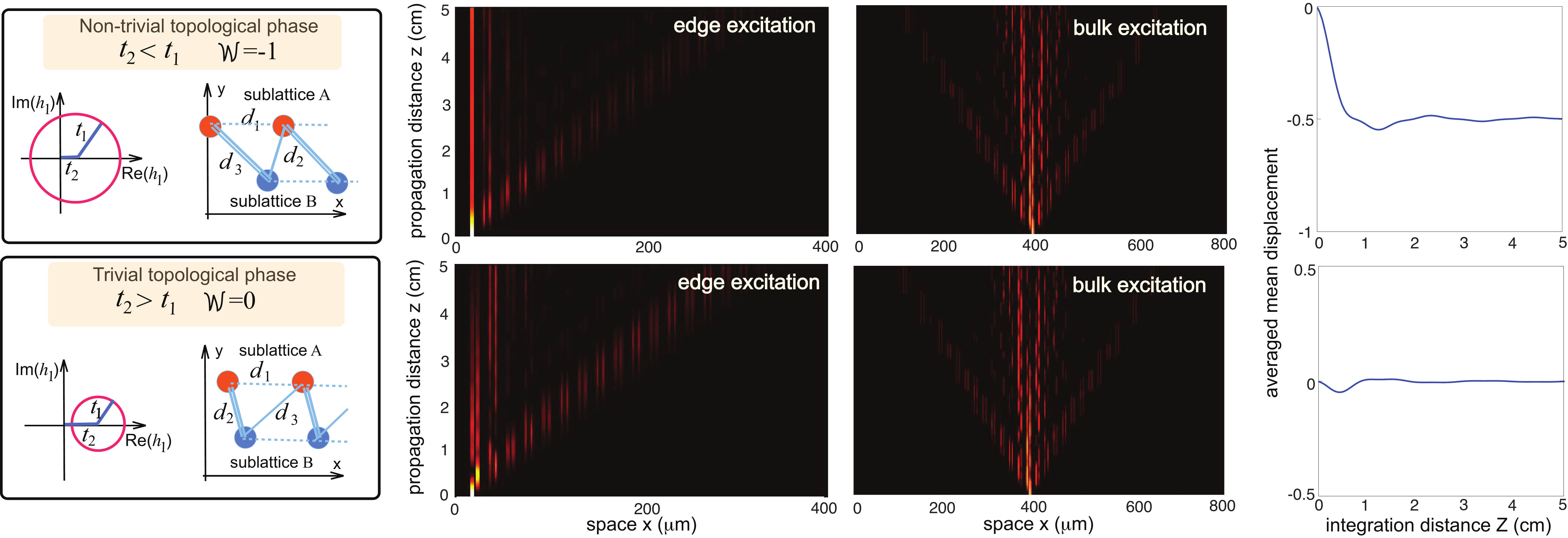}} \caption{ \small
(Color online) Photonic realization of the extended SSH model with broken chiral symmetry in the non-trivial (upper panels) and trivial (lower panels) topological phases. Left column: waveguide array geometry in the transverse $(x,y)$ plane, and rotation of vector $h_1(k)$ as $k$ spans the Brillouin zone. The winding number $\mathcal{W}$ is -1 (0) in the non-trivial (trivial) topological phase. The second and third column panels show the numerically-computed discrete diffraction patterns along the propagation distance $z$ for edge and bulk single waveguide excitation of the sublattice A. Bright (dark) regions correspond to high (low) beam intensity. The far right panels in the figure show the averaged mean spatial displacement of the optical beam versus integration distance $Z$. Parameter values used in the numerical simulations are given in the text.}
\end{figure*}
for a sufficient long observation time $T$. A comparison of Eqs.(8) and (11) shows that, in general, the averaged wave packet displacement does not reproduce any of the Zak phases of the two lattice bands. However, when the Hamiltonian has inversion symmetry and the Zak phase is quantized, i.e. for $\theta(k)= \pi/2$, one obtains  $\overline{\langle n(t) \rangle} \simeq \mathcal{W}/2$.\par
Photons propagating along a binary waveguide lattice can provide a simple and experimentally accessible testbed of topological nontrivial properties of 1D Hamiltonians with broken chiral symmetry. A schematic of the  waveguide lattice is shown in Fig.1(a) and consists of a binary array in a zig-zag geometry. The lattice emulates the extended SSH Hamiltonian \cite{r27,r28} defined by Eqs.(1) and (3) with $\delta_A=\delta_B=0$ and $t_A=t_B$. In the waveguide lattice, $t_1$, $t_2$ and $t_A=t_B$ are determined by the evanescent mode coupling of adjacent waveguides, and thus the ratios $t_1 / t_2$ and $t_A/t_2$  can be conveniently tailored by proper waveguide spacing design \cite{r29,r30}.  We consider here the non-trivial topological phase $t_2<t_1$, corresponding to a non-vanishing winding number $\mathcal{W}=-1$ and the existence of two edge modes in the gap according to the bulk-boundary correspondence. 
In the usual SSH model, i.e. without long-range hopping, the band gap is direct [Fig.1(b)] and closes when  $t_2$ approaches $t_1$. For a non-vanishing next-nearest neighbor hopping $t_A=t_B \neq 0$, an inspection of Eq.(4) indicates that the band structure is more involved and three distinct cases arise [Fig.1(c-e)]: (i)  for $t_A<t_2/2$, the gap is open and direct [Fig.1(c)]; (ii) for $t_2/2<t_A<t_1/2$ the gap is open and indirect [Fig.1(d)]; (iii) for $t_A>t_1/2$ the gap is closed [Fig.1(d)].\\ 
Figure 2 shows the energy spectrum of a lattice comprising $N=40$ unit cells with open boundary conditions for increasing values of the ratio $t_A/t_2$. For $t_A=0$, chiral symmetry is unbroken and one retrieves the usual SSH Hamiltonian with two topologically-protected edge states at (nearly) zero energy $E_{edge}=0$ (finite size effects lead to a slight splitting of the two energies due to hybridization of edge states \cite{r1}, which is negligible in our case). For a non vanishing value  $t_A$, the chiral symmetry is broken and the zero-energy property of the edge modes is lifted. A direct calculation of the edge state energy gives the simple relation 
\begin{equation}
E_{edge}=- 2 t_2 (t_A / t_1)
\end{equation}
which is valid up to the gap closing value $t_A=t_1/2$, above which edge states become delocalized. Inversion symmetry ensures that left and right edge states are (nearly) degenerate with the same value (12) of energy. In fact, the operator $\mathcal{R}=\sigma_x$ transforms one edge state into the other one, according to the specular symmetry of the finite chain. This ensures some topological protection of the edge states, despite chiral symmetry is $^{\prime}$trivially$^{\prime}$ broken \cite{r24,r26} by long-range hopping.\\
The topological property of the extended SSH Hamiltonian can be probed in the bulk by measuring the averaged space displacement of the discretized beam along the longitudinal propagation distance $z=ct$  
when one waveguide of sublattice A is excited at the input plane. Waveguide lattices manufactured by femtosecond laser writing \cite{r9,r29,r30} can offer a suitable platform for an experimental observation of the predicted effects. 
As an example, Fig.3 shows numerically-computed discrete diffraction patterns and corresponding averaged beam displacements -- for the non-trivial ($t_2<t_1$, $\mathcal{W}=-1$) and trivial ($t_2>t_1$, $\mathcal{W}=0$) topological phases-- that one would observe in waveguide arrays manufactured in fused silica \cite{r29,r30}. The geometrical setting of waveguides is schematically depicted in the left panels of Fig.3. For the non-trivial topological phase, waveguide distances are set at the values $d_1= 19 \; \mu$m, $d_2= 11 \; \mu$m and $d_3= 16 \; \mu$m, corresponding to coupling constants $t_A=t_B \simeq 1 \; {\rm cm}^{-1}$,  $t_1 \simeq 3 \; {\rm cm}^{-1}$ and $t_2 \simeq 1.5 \; {\rm cm}^{-1}$ at the probing wavelength $\lambda=633$ nm (He-Ne laser) \cite{r30}. The trivial topological phase is simply obtained by flipping distances $d_2$ and $d_3$. The $L=5$-cm-long waveguide array comprises a total number of $2N=80$ waveguides, and is excited in a single waveguide of sublattice A either at the left edge (second column panels of Fig.3) or in the bulk (third column panels in Fig.3). Single waveguide excitation, either at the edge or in the bulk, is routinely achieved by a focusing microscope objective \cite{r9,r29,r30}. Light spreading in the wave-guide lattice along the propagation distance results is a discrete intensity pattern distribution (discrete diffraction pattern), from which the mean transverse beam displacement  $\overline{\langle n(z) \rangle}$ versus integration distance $Z=cT$ can be retrieved \cite{r9}. The displacements are computed in units of the lattice period $d_1$. Clearly, in the trivial topological phase the averaged mean displacement vanishes and there are not edge states (lower panels in Fig.3), whereas in the non-trivial topological phase the averaged mean displacement rapidly converges to $\mathcal{W}/2=-0.5$ and one edge state is clearly visible (upper panels in Fig.3). Therefore, according to the theoretical analysis probing the bulk of the array can predict the topological phase of the extended SSH model with broken chiral symmetry.\par
{\it Non-Hermitian models.} 
As shown in the previous analysis, quantized Zak phases and nontrivial topological bands without chiral symmetry are possible for Hermitian lattices with long-range hopping. Another interesting example of a two-band lattice with quantized Zak phase and  broken (Hermitian) chiral symmetry is provided by the parity-time ($\mathcal{PT}$) symmetric extension of the SSH model, which has been investigated  in some recent works \cite{r9,r10,r14,r14bis,r17,r23}. The Hamiltonian is given by Eqs.(1) and (3) with $t_A=t_B=0$ and $\delta_A=i \gamma$, $\delta_B=-i \gamma$, where $\gamma$ is the balanced gain/loss rate in sublattices A and B \cite{r14}. The gain/loss terms break in a non-trivial way the chiral symmetry since all components $a_{x,y,z}$ do not vanish.  In the unbroken $\mathcal{PT}$ phase, i.e. for $\gamma<|t_1-t_2|$, the real part of the complex Zak phase is quantized and takes the values $0,-\pi$ for $t_2>t_1$ and $t_2<t_1$, respectively \cite{r14bis}. Edge states in the non-trivial topological phase are protected by particle-hole symmetry, rather than (Hermitian) chiral symmetry \cite{r14}. Can the averaged beam displacement method reproduce the Zak phase for this model? The answer is negative. In fact, a simple extension of the previous analysis to the $\mathcal{PT}$-symmetric Hamiltonian indicates that $\overline{ \langle n(t) \rangle}$ is given by Eq.(11), with $\sin^2 \theta(k)=H^2(k)/[H^2(k)-\gamma^2]$ and $H(k)=|t_2+t_1 \exp(-ik)|$. Hence the averaged mean displacement does not yield the correct value of the Zak phase.  
 
{\it Conclusions.}
 To conclude, one-dimensional two-band lattices without chiral symmetry can show non-trivial topological phases, signaled by quantization of the Zak phase, when  inversion symmetry remains unbroken \cite{r26}. Here we have shown that beam displacement in a continuous-time photonic quantum walk, realized in waveguide arrays, can provide a simple method to probe the topological phases in the bulk of Hamiltonians with broken chiral symmetry. As an example, a zig-zag binary waveguide array has been proposed to test the topological properties of the extended SSH Hamiltonian, where next-nearest neighbor couplings break chiral symmetry but leave unbroken the inversion symmetry. A question that remains open is to find other simple methods to test the topological phases in non-Hermitian models, where fractional winding number could be observed and the bulk-edge correspondence is a subtle matter \cite{r31}.


\begin{thebibliography}{99}
\bibitem{r1}
 J.K. Asb\'oth, L. Oroszl\'any,and A. P\'alyi, {\it A Short Course on Topological Insulators}, Lect. Notes Phys. {\bf 919}, (2016).
 \bibitem{r2}
 L. Lu, J.D. Joannopoulos, and M. Solja\v{c}ic,  Nat. Photon. {\bf 8}, 821-829 (2014).
 \bibitem{r3}
T. Ozawa, H.M. Price, A. Amo, N. Goldman, M. Hafezi, L. Lu, M. Rechtsman, D. Schuster, J. Simon, O. Zilberberg, and I. Carusotto,  arXiv:1802.04173 (2018).
\bibitem{r3bis}
T. Kitagawa, M.S. Rudner, E. Berg, and E. Demler, Phys. Rev. A {\bf 82}, 033429 (2010). 
\bibitem{r4}
T. Kitagawa, M.A. Broome, A. Fedrizzi, M.S. Rudner, E. Berg, I. Kassal, A. Aspuru-Guzik, E. Demler, and A.G. White, Nat. Commun. {\bf 3}, 882 (2012).
\bibitem{r5}
Y. E. Kraus, Y. Lahini, Z. Ringel, M. Verbin, and O. Zilberberg, Phys. Rev. Lett. {\bf 109}, 106402 (2012).
\bibitem{r6}
M. Hafezi, S. Mittal, J. Fan, A. Migdall, and J.M. Taylor, Nat. Photon. {\bf 7}, 1001-1005 (2013).
\bibitem{r7}
M. Hafezi, Phys. Rev. Lett. {\bf 112}, 210405 (2014).
\bibitem{r8}
Y. Plotnik, M.C. Rechtsman, D. Song, M. Heinrich, J.M. Zeuner, S. Nolte, Y. Lumer, N. Malkova, J. Xu, A. Szameit, Z. Chen, and M. Segev, Nat. Mat. {\bf 13}, 57-62 (2014).
\bibitem{r9} 
J.M. Zeuner, M.C. Rechtsman, Y. Plotnik, Y. Lumer, S. Nolte, M.S. Rudner, M. Segev, and A. Szameit, Phys. Rev. Lett. {\bf 115}, 040402 (2015).
\bibitem{r10}
C. Poli, M. Bellec, U. Kuhl, F. Mortessagne, and H. Schomerus, Nat. Commun. {\bf 6}, 6710 (2015).
\bibitem{r11}
F. Cardano, M. Maffei, F. Massa, B. Piccirillo, C. de Lisio, G. De Filippis, V. Cataudella, E. Santamato, and L. Marrucci, Nat. Commun. {\bf 7}, 11439 (2016).
\bibitem{r12}
S. Barkhofen, T. Nitsche, F. Elster, L. Lorz, A. Gabris, I. Jex, and C. Silberhorn, Phys. Rev. A {\bf 96},  033846 (2017).
\bibitem{r13}
F. Cardano, A. D'Errico, A. Dauphin, M. Maffei, B. Piccirillo, C. de Lisio, G. De Filippis, V. Cataudella, E. Santamato, L. Marrucci, M. Lewenstein, and P. Massignan, Nat. Commun. {\bf 8}, 15516 (2017).
\bibitem{r14}
H. Schomerus, Opt. Lett. {\bf 38}, 1912-1914 (2013).
\bibitem{r14bis}
H. Zhao, S. Longhi, and L. Feng, Sci. Rep. {\bf 5}, 17022 (2015).
\bibitem{r15}
L. Xiao, X. Zhan, Z. H. Bian, K. K.Wang, X. Zhang, X. P.Wang, J. Li, K. Mochizuki, D. Kim, N. Kawakami, W. Yi, H. Obuse, B. C. Sanders, and P. Xue, Nat. Phys. {\bf 13}, 1117-1123 (2017).
\bibitem{r16}
S. Weimann, M. Kremer, Y. Plotnik, Y. Lumer, S. Nolte, K. G. Makris, M. Segev, M. C. Rechtsman, and A. Szameit, Nat. Mat. {\bf 16}, 433-438 (2017).
\bibitem{r16bis}
D. Leykam, Daniel, K. Y. Bliokh, C. Huang, Y.D. Chong, and F. Nori, Phys. Rev. Lett. {\bf 118}, 040401 (2017)
\bibitem{r17}
M. Pan, H. Zhao, P. Miao, S. Longhi, and L. Feng, Nat. Commun. {\bf 9}, 1308 (2018).
\bibitem{r19}
J. Zak, Phys. Rev. Lett. {\bf 62}, 2747-2750 (1989).
\bibitem{r20}
M. Atala, M. Aidelsburger, J.T. Barreiro, D. Abanin, T. Kitagawa, E. Demler, and I. Bloch, Nat. Phys. {\bf 9}, 795-800 (2013).
\bibitem{r21}
S. Longhi, Opt. Lett. {\bf 38}, 3716-3719 (2013).
\bibitem{r22}
M. Cominotti and I. Carusotto, EPL {\bf 103}, 10001 (2013).
\bibitem{r23}
M. S. Rudner and L. S. Levitov, Phys. Rev. Lett. {\bf 102}, 065703 (2009).
\bibitem{uff}
S. Ryu, A.P. Schnyder, A. Furusaki, and A.W.W. Ludwig, New J. Phys. {\bf 12}, 065010 (2010).
\bibitem{r24}
G. van Miert, C. Ortix, and C. Morais Smith, 2D Mater. {\bf 4}, 015023 (2017).
\bibitem{r26}
B. Song, L. Zhang, C. He, T. Fung J. Poon, E. Hajiyev, S. Zhang, X.-J. Liu, and G.-B. Jo, Sci. Adv. {\bf 4}, eaao4748 (2018).
\bibitem{referee}
X. Yin, Z. Ye, J. Rho, Y. Wang, and X. Zhang, Science {\bf 339}, 1405-1407 (2014). 
\bibitem{r27}
L. Li, Z. Xu, and S. Chen,  Phys. Rev. B {\bf 89}, 085111 (2014).
\bibitem{r28}
B. Perez-Gonzalez, M. Belloy, A. Gomez-Leon, and G. Platero, arXiv:1802.03973v1.
\bibitem{r29}
F. Dreisow, A. Szameit, M. Heinrich, T. Pertsch, S. Nolte, and A. T\"unnermann, Opt. Lett. {\bf 33}, 2689-2691 (2008).
\bibitem{r30}
 G. Corrielli, A. Crespi, G. Della Valle, S. Longhi, and R. Osellame, Nat. Commun. {\bf 4}, 1555 (2013).
 \bibitem{r31}
 T.E. Lee, Phys. Rev. Lett. {\bf 116}, 133903 (2016).
 
\end{thebibliography}
\end{document}